\documentclass[aps,twocolumn,superscriptaddress]{revtex4-2}
\usepackage{geometry}
\usepackage{amsfonts,amssymb,amsmath}
\usepackage[colorlinks=true]{hyperref}
\usepackage{graphicx}
\usepackage{color}
\usepackage{xcolor}
\usepackage{dcolumn}
\usepackage{bm}
\usepackage{hyperref}
\usepackage{physics}

\newcommand{\aver}[1]{\left\langle #1 \right\rangle}
\newcommand{\ft}{\widetilde{f}}
\newcommand{\gt}{\widetilde{g}}
\newcommand{\st}{\widetilde{s}}

\newcommand{\Gammat}{{\widetilde{\Gamma}}}
\newcommand{\Gt}{{\widetilde{G}}}

\newcommand{\pval}{\mathrm{p.v.}\!}
\newcommand{\Ht}{{\widetilde{H}}}

\begin{document}

\title{An exactly solvable model of wave-mean field interaction in integrable turbulence}

\author{T. Congy}
\email{thibault.congy@northumbria.ac.uk}
\author{G. A. El}
\affiliation{Department of Mathematics, Physics and Electrical Engineering, Northumbria University, Newcastle upon Tyne NE1 8ST, United Kingdom}
\author{M. A. Hoefer}
\affiliation{Department of Applied Mathematics, University of Colorado, Boulder, Colorado 80309, USA}

\begin{abstract}
The kinetic theory of soliton gases (SG) is used to develop a solvable  model for wave-mean field interaction in integrable turbulence. The waves are stochastic soliton ensembles that scatter off a critically dense SG or soliton condensate---the mean field. The derived two-fluid kinetic-hydrodynamic equations admit exact solutions predicting SG filtering and an induced mean field. The obtained SG statistical moments agree with ensemble averages of numerical simulations. The developed theory readily generalizes, with applications in fluids, nonlinear optics and condensed matter. 
\end{abstract}

\maketitle

At the vanguard of quantum many-body physics \cite{ doyon_generalized_2025}, geophysical fluid dynamics \cite{Costa:14, Redor:19,
suret_nonlinear_2020, Redor:21,fache_interaction_2024}, nonlinear optics \cite{kraych_statistical_2019,marcucci_topological_2019,suret_soliton_2023}, and soliton theory \cite{suret_soliton_2024} is the evolution of random initial states according to an integrable or near-integrable system.  Dubbed integrable turbulence \cite{zakharov_turbulence_2009}, this class of multiscale nonlinear dynamics holds the tantalizing prospect of yielding solvable, physical turbulent models. In contrast, classical fluid turbulence is plagued by the closure problem in which statistical moments of the fluid state depend on successively higher order moments \cite{pope_turbulent_2000}.

The theory of soliton gases (SG) describes random ensembles of many interacting solitons via a kinetic equation for the evolution of the density of states (DOS), the distribution of solitons with respect to their amplitudes and spatial positions \cite{el_soliton_2021}.  Closure is achieved in the thermodynamic limit ($N \to \infty$) of modulated $N$-phase (quasi-periodic) solutions \cite{el_thermodynamic_2003}, the existence of which is intimately connected to an infinite number of independent conserved quantities.  A related strand of research is generalized hydrodynamics (GHD) \cite{doyon_generalized_2025}, in which integrable quantum and classical many-body systems are described by the out-of-equilibrium dynamics of interacting quasi-particles on large spatiotemporal scales.  A correspondence between SG and GHD was established for soliton ensembles in the Korteweg-de Vries (KdV) equation \cite{bonnemain_generalized_2022}.

A related counterpart to integrable turbulence is wave turbulence (WT) modeled by statistical ensembles of interacting small-amplitude dispersive waves \cite{newell_wave_2011,nazarenko_wave_2011}.  While closure is achieved and analytical solutions are known, the theory is restricted to nearly linear, small-amplitude waves and does not account for the formation of localized, finite-amplitude coherent structures like solitons and breathers, which are ubiquitous in strongly nonlinear wave fields. WT, SG and GHD theories also feature large-scale coherent states, or condensates \cite{connaughton_condensation_2005,congy_dispersive_2023,doyon_large-scale_2017} that can be viewed as mean fields.  However, a theory that couples random waves or quasi-particles and a mean field is lacking.

\begin{figure*}
    \centering   
    \includegraphics[scale=0.1]{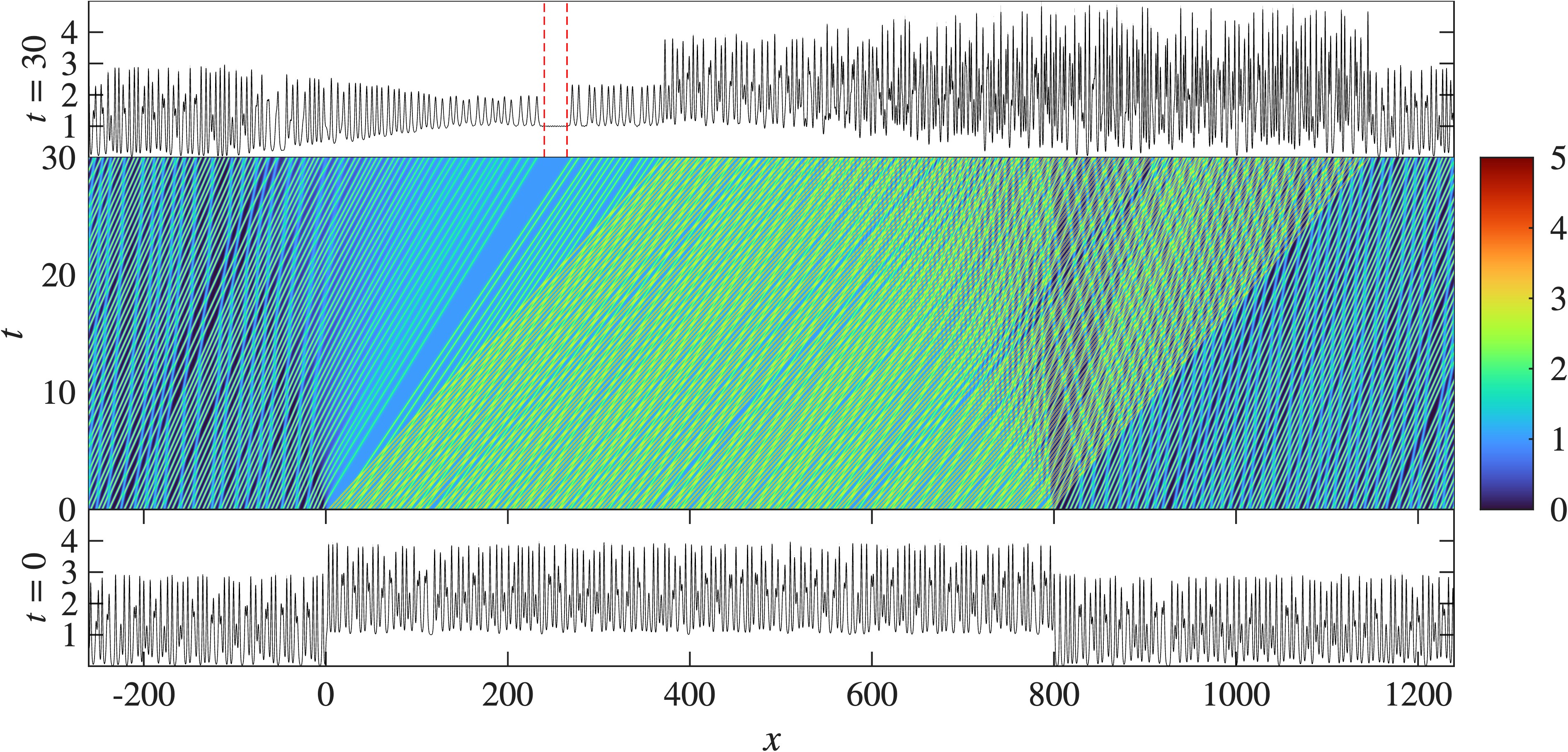}
    \caption{Wave-mean field interaction in KdV soliton gas (SG) integrable turbulence. Lower panel: KdV initial condition for a wide, localized unit mean-field elevation
added to a    two-component SG with DOS $f(\eta) = 0.08 \delta(\eta-0.8)+0.17 \delta(\eta-1.2)$ for 400 solitons.
    Middle panel:  contour plot of initial condition evolution featuring SG interaction with stochastic rarefaction and dispersive shock waves.  Upper panel: the turbulent wave field at $t = 30$.}
    \label{fig:example}
\end{figure*}

In this Letter, we develop a two-fluid theory of SG for wave-mean field interaction (see a numerical realization in Fig.~\ref{fig:example}), a long-standing problem in geophysics \cite{buhler_waves_2014}.    
We derive fully coupled kinetic-hydrodynamic equations for the KdV soliton DOS---describing the stochastic SG, an analogue of the normal fluid in the two-fluid theory of superfluidity \cite{landau_theory_1941}---and a mean-field or soliton condensate \cite{congy_dispersive_2023}---describing the deterministic soliton superfluid.  A key finding is the existence of a soliton induced mean-field.
The  SG-mean field equations are diagonalized and solutions are obtained. For a polychromatic SG \cite{congy_riemann_2025} scattering off of a dynamic mean-field as in Fig.~\ref{fig:example}, the theory predicts SG filtering, the spontaneous creation of pure superfluid states, stochastic rarefaction and dispersive shock waves, and contact shocks. The developed theory is general and applies to other integrable systems with applications in nonlinear physics.

The theory of SG utilizes the inverse scattering transform that maps discrete spectral points $\zeta_j$ of the associated Lax operator to solitons \cite{ablowitz1981solitons, novikov1984theory}. Here, we focus on the KdV equation $u_t + 6uu_x+u_{xxx}=0$ whose Lax operator is $-\partial^2_{xx} - u$.  We transform its spectrum $\zeta_j = - \eta_j^2$, so that $\eta_j \in \mathbb{R}^+$ and the soliton's amplitude is $a_j = 2 \eta_j^2$. The DOS $f(\eta;x,t)$ (we sometimes suppress the $(x,t)$ dependence for simplicity) is the joint distribution of solitons with respect to their spectral parameters and positions; it satisfies \cite{el_soliton_2021}
\begin{equation} \label{eq:kinetic} 
f_t + (sf)_x = 0,  
\end{equation}
where $s(\eta;x,t)$ is the effective soliton velocity
\begin{equation}\label{eq:state}
s(\eta) = s_0(\eta) + \int_\Gamma G(\eta, \mu) f(\mu)[s(\eta)-s(\mu)] d\mu .
\end{equation}
Here, $s_0$ is the free soliton velocity, $G$ the two-soliton scattering phase-shift kernel, and $\Gamma \subset \mathbb{R}^+$ is the DOS spectral support. For KdV, 
\begin{equation}\label{kdv_ingr}
s_0(\eta) = 4\eta^2,\ \ G(\eta, \mu)= \frac 1 \eta \ln \left|\frac{\eta+\mu}{\eta-\mu} \right|.
\end{equation}  
The ensemble averages of the physical wave field can be expressed in terms of DOS moments, e.g., for KdV  \cite{el_thermodynamic_2003,   el_soliton_2021}
\begin{equation}\label{aver}
  \aver{u} =\int_\Gamma 4\eta f(\eta) d\eta, \ \ 
  \aver{u^2} =\int_\Gamma \frac{16}{3} \eta^3 f(\eta) d\eta .
\end{equation}

Originally derived for KdV dilute \cite{zakharov_kinetic_1971} and dense \cite{el_thermodynamic_2003} SG, the kinetic equation \eqref{eq:kinetic} universally describes a weakly inhomogeneous SG in integrable systems \cite{el_soliton_2021,suret_soliton_2024}. Independently, the same equation arose in GHD \cite{doyon_lecture_2020,spohn_hydrodynamic_2023,doyon_generalized_2025}.

It was shown in \cite{congy_dispersive_2023} that coherent dispersive hydrodynamic mean fields such as a rarefaction or dispersive shock wave \cite{El2016Dispersive} represent non-equilibrium {\it soliton condensates}, critically dense SG 
specified by certain nonlinear dispersion relations. Soliton condensates are characterized by a non-negative integer $N$ called the genus that is the number of gaps in the spectral support denoted $\Gamma_N$.
Generally, SG include two components that form a {\it composite SG} \cite{biondini_breather_2025}: a condensate, modeling a coherent mean field or superfluid, and a non-condensate, modeling a stochastic SG or normal fluid. Then, wave-mean field interaction in integrable turbulence and a two-fluid theory of SG is described by their coupling.  Our primary contribution in this Letter is a suitable partitioning of the DOS $f(\eta)$ and its spectral support $\Gamma$ that realizes this two-fluid theory of wave-mean field interaction.

Introducing the DOS spectral support 
\begin{equation}
\label{eq:support}
\Gamma = \Gamma_N \cup \tilde \Gamma, \quad \Gamma_N \cap \tilde \Gamma = \emptyset,
\end{equation}
then the composite SG DOS $f(\eta)$ is
\begin{equation}
\label{eq:composite}
f(\eta) = \underbrace{f^{(N)}(\eta) - f_{\text{IMF}}(\eta)}_{\mathrm{condensate}} \hspace{2mm}+ \hspace{-5mm}\underbrace{\ft(\eta)}_{\mathrm{non-condensate}}
\end{equation}
where the \textit{core} condensate DOS $f^{(N)}(\eta)$ and the {\it induced mean field} condensate DOS $f_\text{IMF}(\eta)$ are both supported on $\Gamma_N$, while the non-condensate DOS $\ft(\eta)$ is supported on $\Gammat$,  equaling zero at endpoints of $\Gammat$.

Using eqs.~\eqref{eq:support} and \eqref{eq:composite} in the kinetic eq.~\eqref{eq:kinetic} and equation of state \eqref{eq:state}, we derive a general two-fluid theory for a composite SG corresponding to the interaction of a KdV SG with a genus $N$ soliton condensate in the End Matter where $f^{(N)}$ and $f_{\rm IMF}$ are defined by eqs.~\eqref{eq:condensate_NDR}, \eqref{eq:ansatz}, \eqref{eq:induced} and $\ft$ satisfies eqs.~\eqref{eq:kint}, \eqref{eq:statet}, \eqref{eq:dressed_s0_G}. Here, we highlight the simplest case with the  genus 0 core soliton condensate DOS \cite{congy_dispersive_2023}
\begin{equation}
    f^{(0)}(\eta) = \frac{\eta}{\pi\sqrt{\beta^2-\eta^2}},\quad \Gamma_0=[0,\beta).
\end{equation}
If $\ft \equiv 0$, the slow modulation of the condensate is enabled by allowing $\Gamma_0 = \Gamma_0(x,t)$ to evolve
\begin{equation}
\label{eq:beta0}
    \beta_t + 6\beta^2 \beta_x = 0, \quad \ft(\eta) \equiv 0.
\end{equation}
More generally, if $\tilde f \ne 0$ the induced mean field is
\begin{equation}
\label{eq:imf0}
\begin{split}
&f_{\text{IMF}}(\eta) = \int_\beta^\infty p^{(0)}(\eta,\mu) \ft(\mu)d\mu,\\
&p^{(0)}(\eta,\mu) = \frac{2\eta \sqrt{\mu^2-\beta^2}}{\pi(\mu^2-\eta^2)\sqrt{\beta^2-\eta^2}},
\end{split}
\end{equation}
where $\Gammat = [\beta,\infty)$.

Letting $\ft = \ft(\eta;x,t)$,
the kinetic equation for the non-condensate component of the composite SG is
\begin{equation}
\label{eq:kint}
    \ft_t+(\st \ft )_x=0,
\end{equation}
where  $\st(\eta)$ satisfies the equation of state
\begin{equation}
\label{eq:statet}
    \st(\eta) = \st_0(\eta) + \int_\beta^\infty \!\! \Gt(\eta,\mu) \ft(\mu)[\st(\eta)-\st(\mu)] d\mu,
\end{equation}
\begin{equation}
\label{eq:dressedcoeff}
\begin{split}
    &\st_0(\eta) = 2\beta^2+4\eta^2,\\ &\Gt(\eta,\mu) =G\Big(\sqrt{\eta^2-\beta^2},\sqrt{\mu^2-\beta^2}\Big).
\end{split}
\end{equation}
The soliton free velocity $\st_0$ and the two-soliton scattering shift $\Gt$ are modified or ``dressed'' by the coupling to the condensate.

The condensate component in the composite SG satisfies the modulation equation
\begin{equation}
\label{eq:Whitham}
   \beta_{t} + \gamma \beta_{x} = 0,\;\; \gamma= \frac{ 6\beta^2 - \int_\beta^\infty \frac{2\st(\mu) \ft(\mu)}{\sqrt{\mu^2-\beta^2}}d\mu}{ 1 - \int_\beta^\infty \frac{2\ft(\mu)}{\sqrt{\mu^2-\beta^2}}d\mu}.
\end{equation}

Equations \eqref{eq:support}--\eqref{eq:Whitham} comprise our main result:  the  kinetic-hydrodynamic two-fluid model for the DOS $f(\eta;x,t)$ and spectral support $\Gamma(x,t)$ of wave-mean field interaction in a composite SG.   See the End Matter for the more general case of genus $N \ge 0$.

Substituting \eqref{eq:support}--\eqref{eq:imf0} in \eqref{aver} yields
\begin{equation}
\label{eq:aver0}
\begin{split}
  &\aver{u} = \beta^2 + \int_\beta^\infty 4\sqrt{\eta^2-\beta^2} \ft(\eta) d\eta, \\
  &\aver{u^2} = \beta^4 + \int_\beta^\infty \frac83 (2\eta^2+\beta^2)\sqrt{\eta^2-\beta^2} \ft(\eta) d\eta.
  \end{split}
\end{equation}
In the absence of a non-condensate SG ($\ft \equiv 0$), the mean field is $\aver{u} = \beta^2$ and the variance $\mathrm{Var}(u) = \aver{u^2}-\aver{u}^2 = 0$, i.e., the flow \eqref{eq:beta0} corresponds to a pure, deterministic superfluid for the mean field $\beta^2$.  The addition of a non-zero, non-condensate SG in eq.~\eqref{eq:aver0} results in positive variance and a positive induced mean field since $\aver{u} > \beta^2$.  

We now use these results to describe the interaction of a dense KdV SG with a genus zero soliton condensate in a stochastic rarefaction wave initiated by step Riemann data \cite{congy_dispersive_2023}. This problem generalizes soliton-mean field interaction \cite{maiden_solitonic_2018, ablowitz_solitonmean_2023}, and its accompanying soliton transmission through or trapping inside the mean field, to the complex nonlinear interactions of a dense SG with a mean field as in Fig.~\ref{fig:example}.

For analytical results, we focus on a multi-component (polychromatic) SG for the non-condensate DOS
\begin{equation}\label{poly}
\ft(\eta;x,t)= \sum \limits_{j=1}^M w_j(x,t) \delta(\eta - \xi_j),
\end{equation}
where $\xi_j \in \Gammat$, $j=1,\ldots,M$.  The delta-function ansatz \eqref{poly} is a mathematical idealization of a narrow spectral distribution around $\xi_j$, valid for $\xi_j$ sufficiently separated from $\beta$.  In this case, the kinetic-hydrodynamic system  \eqref{eq:kint}--\eqref{eq:Whitham} reduces to the hydrodynamic conservation laws 
\begin{equation}\label{hyd_cons}
\begin{split}
& w_{j,t}+(c_j\,w_j)_x=0, \ \ j=1, \dots, M, \\ 
&c_j = 2\beta^2+ 4\xi_j^2 +  \sum_{k\neq j} \Gt_{jk} w_k(c_j-c_k), 
\end{split}
\end{equation}
with $\Gt_{jk} =\Gt (\xi_j, \xi_k)$ that is
coupled to the mean field evolution
\begin{equation}
\label{hyd_cons_la}
 \beta_{t} + \gamma \beta_{x} = 0,\;\; \gamma = \frac{ 6\beta^2 - \sum_{j=1}^M \frac{2 c_jw_j}{\sqrt{\xi_j^2-\beta^2}}}{1-\sum_{j=1}^M \frac{2w_j}{\sqrt{\xi_j^2-\beta^2}}}.
\end{equation}

The combined system \eqref{hyd_cons}, \eqref{hyd_cons_la} exhibits important mathematical properties with physical implications.  The system \eqref{hyd_cons} with $\beta \equiv 0$ has been shown to be a linearly degenerate, integrable, hyperbolic system of hydrodynamic type \cite{el_kinetic_2005,el_kinetic_2011}.
Equation \eqref{hyd_cons} with $\beta$ satisfying \eqref{hyd_cons_la} remains linearly degenerate \cite{supplemental}, implying the absence of wave breaking and that only contact shock discontinuities moving with characteristic speed $c_j$ are admissible. Equation \eqref{hyd_cons_la} for the mean flow is not strictly genuinely nonlinear $\partial_{\beta}\gamma  \not\equiv 0$ \cite{lax_hyperbolic_1973}, which affects solutions of the Riemann problem.  The system \eqref{hyd_cons} can be diagonalized \cite{supplemental}
\begin{equation}
\label{eq:qj}
\begin{split}
    &q_{j,t}+c_j q_{j,x}=0, \ \ j=1, \dots, M, \\ 
    &q_j =  (1-\sum_{k\neq j} \Gt_{jk} w_k ) \sqrt{\xi_j^2-\beta^2}/w_j\\
    &\qquad - \ln ( 1-\beta^2/\xi_j^2 ),
\end{split}
\end{equation}
so that the coupled system for $\{\beta\} \cup \{q_j\}_{j=1}^M$ is diagonal and hyperbolic. We prove in \cite{supplemental} that the system \eqref{hyd_cons}, \eqref{hyd_cons_la} is integrable via the generalized hodograph transform \cite{tsarev_poisson_1985,tsarev_geometry_1991} for $M=2$.

\begin{figure}
    \centering
    \includegraphics[scale=0.25]{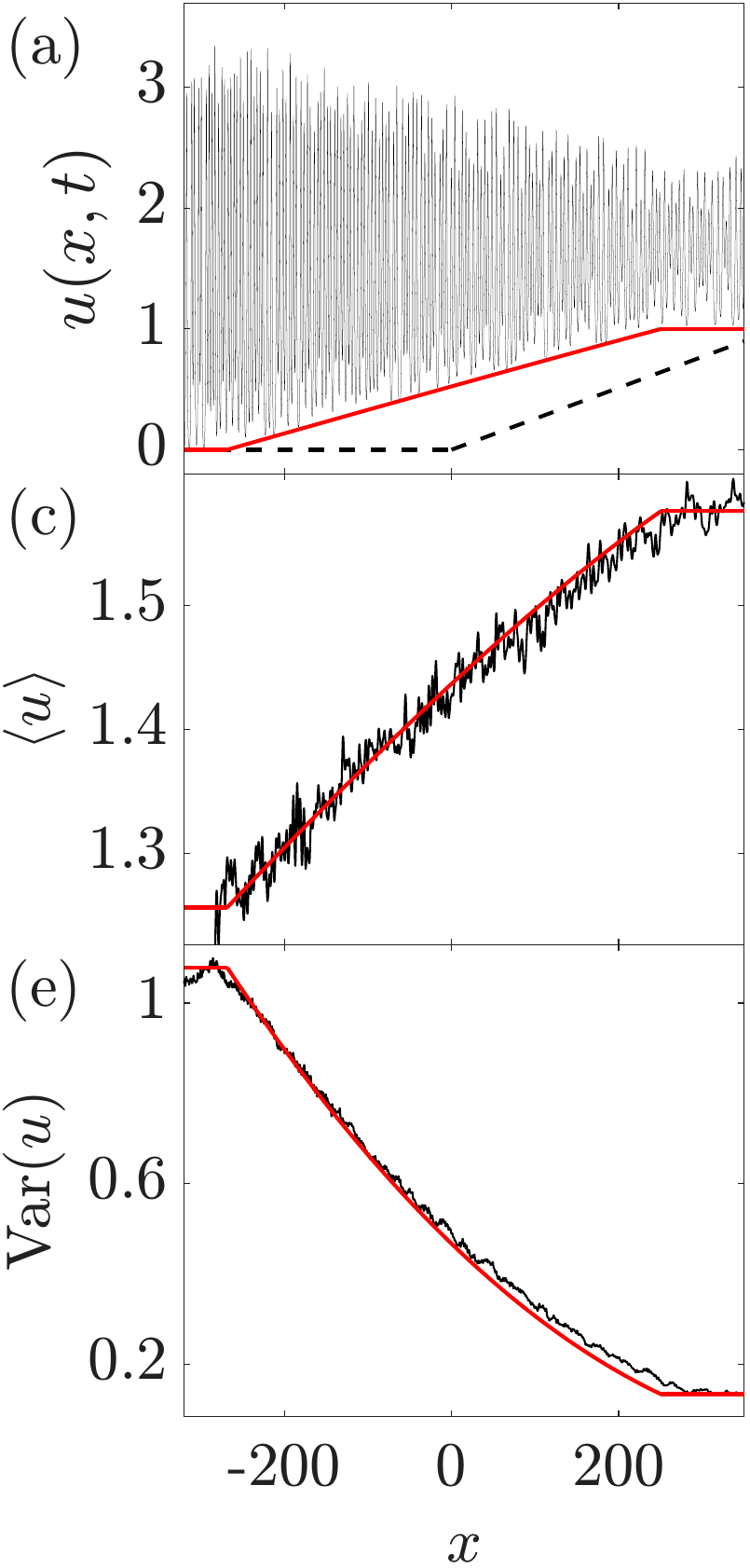}
    \includegraphics[scale=0.25]{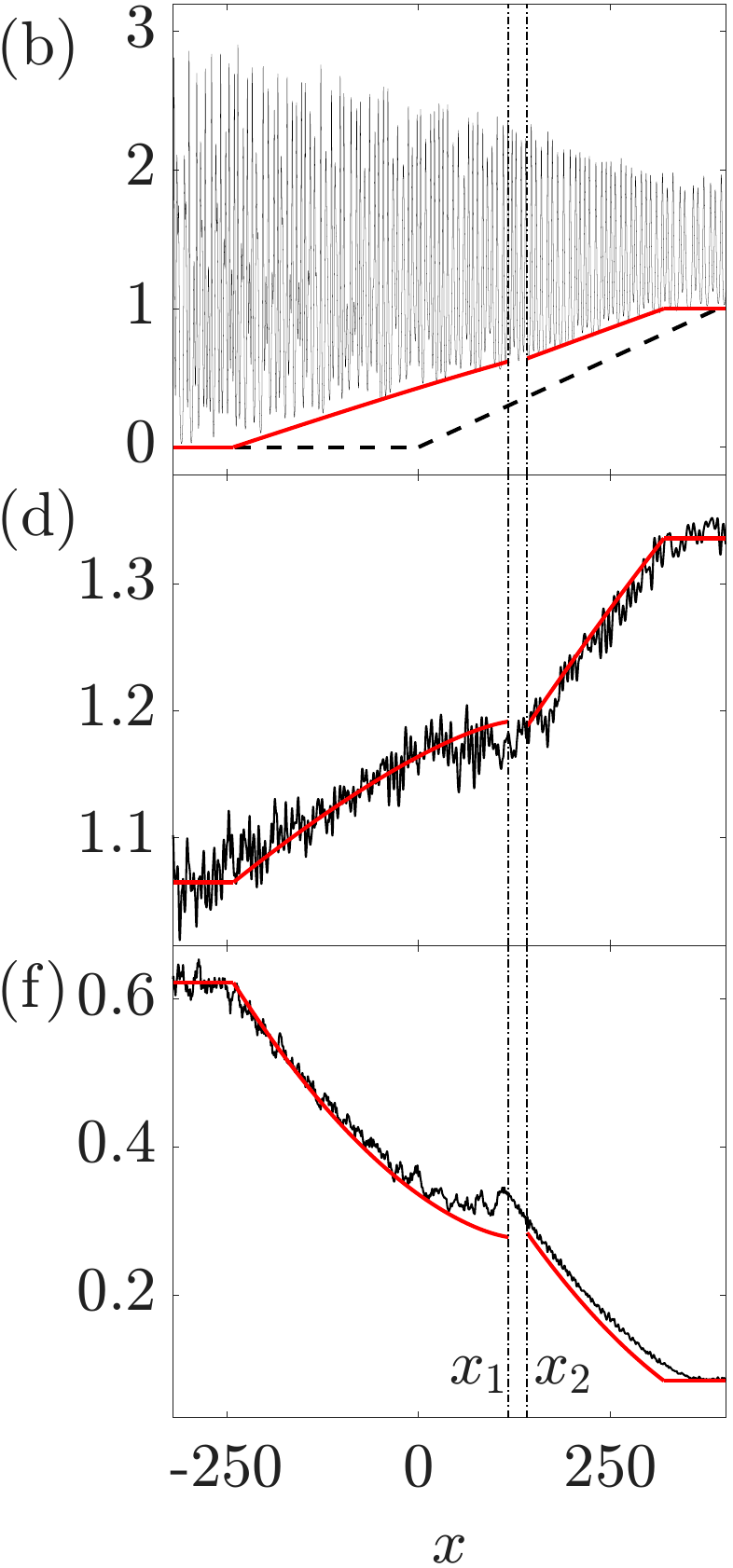}
    \caption{SG interaction with a mean-field rarefaction wave (left half of Fig~\ref{fig:example}). (a,b) Bi-chromatic SG transmission with $(\xi_1,\xi_2)=(1.2,1.3)$ and $(w_{10},w_{20})=(0.1,0.14)$. (c,d) Bi-chromatic SG filtering with $(\xi_1,\xi_2)=(0.8,1.2)$ and $(w_{10},w_{20})=(0.08,0.17)$. (a,c) One realization of the soliton gas at $t=65$. (b,d) Comparison between the analytical results \eqref{eq:aver0} and ensemble averages of $100$ realizations.  See main text for additional details.}
    \label{fig:results}
\end{figure}

Prompted by the simulation in Fig.~\ref{fig:example}, we consider a 2-component (bi-chromatic) non-condensate SG with densities $w_1(x,t)$, $w_2(x,t)$ satisfying eq.~\eqref{hyd_cons} and a genus 0 soliton condensate $\beta(x,t)$ with the step initial condition 
\begin{equation}
\label{eq:init}
    (\beta,w_1,w_2)\big |_{t=0} = \begin{cases}
        (0,w_{1}^0,w_{2}^0), &x<0,\\
        (1,0,0), &x>0.
    \end{cases}
\end{equation}
This represents the collision of a pure normal fluid for $x < 0$ with a pure superfluid for $x > 0$.
The solution of this Riemann problem describes the wavefield to the left of the vertical dashed red lines in Fig.~\ref{fig:example}.  Figure \ref{fig:results}(a,b) display representative numerical realizations for $t\gg 1$ of two distinct scenarios in which $1 < \xi_1 < \xi_2$ and $0 < \xi_1 < 1 < \xi_2$, respectively. Details of the numerical method are given in \cite{supplemental}. A visual examination reveals that the bi-chromatic SG penetrates through the mean field rarefaction wave to emerge on the unit background in Fig.~\ref{fig:results}(a), i.e., the SG is transmitted.  Conversely, in Fig.~\ref{fig:results}(b), the mean field acts as a SG filter allowing one component of the bi-chromatic SG to transmit while the other remains trapped inside the varying mean.

We compare the analytical solution to the Riemann problem \eqref{eq:init} (see End Matter for details) with an ensemble of 100 numerical realizations of the SG.  Figure \ref{fig:results}(c--f) depicts the mean and variance computed from ensemble averages (black curves) and eq.~\eqref{eq:aver0} (red curves).  In the interval $x\in[x_1,x_2]$ denoted by vertical dashed lines, $\beta(x,t)$ and $\xi_1$ are no longer well-separated, breaking the delta function ansatz \eqref{poly}.  This leads to the local loss of genuine nonlinearity $\partial_\beta \gamma = 0$ and a singularity in $\gamma$.  It can be resolved by replacing $\delta(\eta-\xi_1)$ in \eqref{poly} with a smooth, narrow distribution centered at $\xi_1$ and solving \eqref{eq:kint}--\eqref{eq:Whitham}.

Returning to Fig.~\ref{fig:results}(a,b), the rarefaction wave solution $\beta^2$ of the mean field equation \eqref{eq:beta0} without the stochastic SG component $(w_1^0,w_2^0)=(0,0)$ is plotted with a dashed curve. Due to the coupling of the composite SG when $(w_1^0,w_2^0) \ne (0,0)$, the solitons constituting the condensate are phase-shifted leftward owing to their interaction with the solitons in the bi-chromatic SG.  This results in an overall backflow of the condensate as displayed by the solution of \eqref{hyd_cons_la} plotted as the solid red curve.

In conclusion, we have developed a two-fluid theory of composite soliton gases (SG) in the KdV equation consisting of deterministic condensate and stochastic non-condensate components.  The condensate corresponds to a mean field, analogous to a superfluid, whereas the non-condensate is analogous to a normal fluid.  In contrast to deterministic soliton steering by the mean field, a one-way interaction in soliton-mean field theory \cite{maiden_solitonic_2018,ablowitz_solitonmean_2023}, both condensate and non-condensate components influence one another, leading to an induced mean field and non-trivial interactions such as SG filtering by the mean.  Another effect is the spontaneous emergence of a zero variance state or pure superfluid due to the slowing down of the SG by the mean, identified in Fig.~\ref{fig:example} by the vertical dashed lines.  We obtained the kinetic solution and statistics for a stochastic rarefaction wave describing Fig.~\ref{fig:example} for $x<200$.  For $x > 200$, the initial condition is a step down in the mean that, in the absence of a stochastic SG, gives rise to a dispersive shock wave (DSW).  An important extension of this work will be to stochastic DSWs. Also, the derived kinetic-hydrodynamic system and its hydrodynamic reductions represent a new class of integrable systems worthy of further investigation.  Importantly, the approach taken here can be generalized to other integrable systems such as the nonlinear Schr\"odinger equation with applications in condensed matter, optical, and nonlinear physics.

\begin{acknowledgments}
We thank the Isaac Newton Institute for Mathematical
Sciences, Cambridge, for support and hospitality during
the program Emergent Phenomena in Nonlinear Dispersive
Waves, where work on this paper was initiated. This work
was supported by EPSRC Grant No. EP/V521929/1.
 The work of M.H. was supported by NSF grant DMS-2306319 and the Leverhulme  Trust Visiting Professorship at Northumbria University, UK.
\end{acknowledgments}


\section*{End Matter}

{\bf Two-fluid kinetic description:} The DOS
$f^{(N)}$ and the corresponding spectral flux density $g^{(N)}$ are defined by the nonlinear dispersion relations (NDRs)
 \begin{equation}
    \label{eq:condensate_NDR}
    \begin{split}
           & \int_{\Gamma_N}\!\! G(\eta,\mu) f^{(N)}(\mu) d\mu = 1,~ \eta \in \Gamma_N, \\
           &\int_{\Gamma_N}\!\! G(\eta,\mu) g^{(N)}(\mu) d\mu = s_0(\eta),~ \eta \in \Gamma_N,
    \end{split}
\end{equation}
with $\Gamma_N = [0,\beta_1) \cup (\beta_2,\beta_3) \dots \cup (\beta_{2N},\beta_{2N+1})$, $N$ being the condensate genus; for simplicity we dropped the index $i$ in $\beta_i$ when $N=0$. Explicit solutions $f^{(N)}(\eta)$, $g^{(N)}(\eta)$ to the condensate NDRs that are parameterized by the $\beta_i$'s were found in \cite{kuijlaars_minimal_2021, congy_dispersive_2023} with the kinetic equation \eqref{eq:kinetic} mapping to the $N$-phase Whitham modulation equations for the spectral band edges \cite{flaschka_multiphase_1980}
\begin{equation}\label{whitham0}
\beta_{i,t} + v_i^{(N)} \beta_{i,x}=0, \ i=1, \dots, 2N+1,
\end{equation}
where $v_i^{(N)} = \partial_{\beta_i} g^{(N)}(\eta) / \partial_{\beta_i} f^{(N)}(\eta)$ is independent from $\eta$ \cite{supplemental,congy_dispersive_2023}. One has in particular $v^{(0)} = 6\beta^2$.

The induced mean field DOS ansatz in \eqref{eq:composite} is
\begin{equation}
\label{eq:ansatz}
f_{\text{IMF}}(\eta) = \int_{\Gammat} p^{(N)}(\eta,\mu) \ft(\mu) d\mu,
\end{equation}
where the kernel $p^{(N)}(\eta,\mu)$ results from
\begin{equation}\label{eq:cond_comp}
    \int_{\Gamma} G(\eta,\mu) f(\mu) d\mu = 1,\quad \eta \in \Gamma_N,
\end{equation}
equivalent to the requirement that the induced mean field only contributes to the condensate component of the composite SG, i.e., the SG remains a condensate for $\eta \in \Gamma_N$ (cf.~eq.~\eqref{eq:condensate_NDR}). 
Substituting \eqref{eq:support}, \eqref{eq:composite}, \eqref{eq:ansatz} in \eqref{eq:cond_comp}, and using \eqref{eq:condensate_NDR} yields the integral equation for $p^{(N)}(\eta,\mu)$
 \begin{equation}
    \label{eq:induced}
    \begin{split}
      &  \int_{\Gamma_N} G(\eta,z) p^{(N)}(z,\mu) dz = G(\eta,\mu), 
        \end{split}
    \end{equation} 
for $(\eta,\mu) \in \Gamma_N \times \Gammat$, which represents the induced mean field NDR additional to \eqref{eq:condensate_NDR}; the solution of \eqref{eq:induced} for $N=0$ is derived in \cite{supplemental}.

We introduce a counterpart of \eqref{eq:composite} for the spectral flux density
\begin{equation}
\label{eq:g}
    g(\eta)  = g^{(N)}(\eta) + \gt(\eta) - \int_{\Gammat} p^{(N)}(\eta,\mu) \gt(\mu) d\mu,
\end{equation}
where $\gt(\eta) = \st(\eta) \ft(\eta)$, and substitute \eqref{eq:support}, \eqref{eq:composite}, \eqref{eq:g} into the kinetic equation \eqref{eq:kinetic}.
Then, restricting to $\eta \in \Gammat$, we obtain the kinetic equation \eqref{eq:kint} for the non-condensate component of the composite SG where  $\st(\eta)$ satisfies the dressed equation of state \eqref{eq:statet} 
(with the substitution $\int_\beta^\infty \to \int_\Gammat$ in the general case), and $\st_0$ and $\Gt$ the free soliton velocity and phase shift kernel modified by the mean field
\begin{equation}
    \label{eq:dressed_s0_G}
    \begin{split}
    &\st_0(\eta)= \frac{s_0(\eta)- \int_{\Gamma_N} G(\eta,\mu) g^{(N)}(\mu) d\mu}{1- \int_{\Gamma_N} G(\eta,\mu) 
    f^{(N)}(\mu) d\mu},\\
    &\Gt(\eta,\mu) = \frac{G(\eta,\mu)- \int_{\Gamma_N} G(\eta,z) p^{(N)}(z,\mu) dz}{1- \int_{\Gamma_N} G(\eta,z) 
    f^{(N)}(z) dz},
    \end{split}
\end{equation}
which simplifies to \eqref{eq:dressedcoeff} for $N=0$.

For $\eta \in \Gamma_N$ in \eqref{eq:kinetic}, the condensate component modified by the induced mean field satisfies the modulation equations (cf.~\eqref{whitham0})
\begin{align}
   & \beta_{i,t} + \gamma_i \beta_{i,x} = 0, \ \ i=1,\dots,2N+1,\\
   &\gamma_i= \frac{v_i^{(N)} - \int_\Gammat a_i^{(N)}(\mu)\gt(\mu) d\mu}{1 - \int_\Gammat a_i^{(N)}(\mu)\ft(\mu) d\mu},
\end{align}
where $
  a_i^{(N)}(\mu) = \partial_{\beta_i} p^{(N)}(\eta,\mu)/\partial_{\beta_i} f^{(N)}(\eta)
$ is independent of $\eta$ \cite{supplemental}. 

\

{\bf Simple wave solution:} 
Suppose first that $1 < \xi_1 < \xi_2$ as in Fig.~\ref{fig:results}(a,b). The stochastic rarefaction wave solution of \eqref{hyd_cons}, \eqref{hyd_cons_la}, \eqref{eq:init} is given implicitly by the simple wave
\begin{equation}
\label{eq:simple1}
\begin{split}
    &\gamma(\beta,w_1,w_2)=x/t,\\ 
    &q_j(\beta,w_1,w_2)= q_j(0,w_1^0,w_2^0),\;j=1,2,
\end{split}
\end{equation}
with $\beta(x,t) \in [0,1]$, $\gamma$ is defined in \eqref{hyd_cons_la} and $q_j$ in \eqref{eq:qj}. Note that this solution is admissible if \eqref{hyd_cons_la} remains genuinely nonlinear ($\partial_\beta \gamma \neq 0$) along the characteristic $\gamma=x/t$, which is the case for the solution \eqref{eq:simple1} plotted in Fig. \ref{fig:results}(b). The transmitted SG emerging to the right of the simple wave is uniform and the corresponding densities $w_j^+$ are determined by the conservation of the two Riemann invariants $q_j(1,w_1^+,w_2^+)= q_j(0,w_1^0,w_2^0)$, $j=1,2$ so that the complete solution is piecewise defined
\begin{equation}
    \label{eq:simple_wave}
    (\beta,w_1,w_2) = 
    \begin{cases}
    (0,w_1^0,w_2^0), & \tfrac{x}{t} \le \gamma^0 , \\
    (\beta,w_1,w_2)(\tfrac{x}{t}), & \gamma^0 < \tfrac{x}{t} < \gamma^+ , \\
    (1,w_1^+,w_2^+), & \gamma^+ \le \tfrac{x}{t} ,
    \end{cases}    
\end{equation}
with $\gamma^0 = \gamma(0,w_1^0,w_2^0)$, $\gamma^+ = \gamma(1,w_1^+,w_2^+)$.
Since $w_j^0 = 0$ in eq.~\eqref{eq:simple1} corresponds to a rarefaction wave, the solution with nonzero $w_j^0$ describes the interaction of a SG with a rarefaction wave, a stochastic rarefaction wave.  
Substituting \eqref{poly}, \eqref{eq:simple1} in \eqref{eq:aver0} yields the modulation of the ensemble average $\langle u \rangle$, $\langle u^2 \rangle$ displayed in 
Fig. \ref{fig:results}(b).

Suppose now that $\xi_1<1<\xi_2$ as in Figs.~\ref{fig:example} and \ref{fig:results}(c,d). 
The modulation of $(\beta,w_1,w_2)$ is given implicitly by the simple wave solution \eqref{eq:simple1}, \eqref{eq:simple_wave} in the genuinely nonlinear region $x < x_1(t)$ where $x_1(t)$ corresponds to the position where $\partial_\beta \gamma=0$, indicated by a vertical dashed line in Fig.~\ref{fig:results}(d). The loss of genuine nonlinearity is due to the logarithmic divergence of $\gamma$ in the region $(x_1,x_2)$ where $\beta$ is in the neighborhood of $\xi_1$, see the explicit formula in \cite{supplemental}. The logarithmic divergence originates from the idealization of the stochastic SG DOS by a sum of delta functions and should be regularized by considering an adequate continuous distribution.
We suppose that the solution \eqref{eq:simple1} is still valid in the region $x>x_2$ where $\beta \in [\xi_1,1]$ and $w_1=0$: the modulation of $(\beta,w_2)$ is then given by the simple wave
\begin{equation}
\begin{split}
    &v(\beta,w_1=0,w_2)=x/t,\\ 
    &q_2(\beta,w_1=0,w_2)= q_2(0,w_1^0,w_2^0),
\end{split}
\end{equation}
for $x_2(t) < x < \gamma(1,0,w_2^+)t$ with $w_2^+$ determined by $q_2(1,0,w_2^+)=q_2(0,w_1^0,w_2^0)$.
This latter assumption is validated numerically by the good agreement between the analytical description and the numerical simulation in the region $x>x_2$ displayed in Fig. \ref{fig:results}(d).
This solution corresponds to SG filtering in which only a portion of the SG is transmitted by the mean field.

%


\clearpage

\onecolumngrid

\begin{center}
\textbf{\large Appendix for: An exactly solvable model of wave-mean field interaction in integrable turbulence}
\end{center}
\setcounter{equation}{0}
\setcounter{figure}{0}
\setcounter{table}{0}
\makeatletter
\renewcommand{\theequation}{S\arabic{equation}}
\renewcommand{\thefigure}{S\arabic{figure}}

The Korteweg-de Vries (KdV) equation is given by:
\begin{equation}\label{eq:kdv}
    u_t + 6uu_x+u_{xxx} = 0.
\end{equation}
The kinetic equation reads:
\begin{equation} \label{eq:kin} 
f_t + g_x = 0, \quad g=sf,
\end{equation}
with the equation of state for the KdV soliton gas (SG):
\begin{equation}\label{eq:statea}
s(\eta) = s_0(\eta) + \int_\Gamma G(\eta, \mu) f(\mu)[s(\eta)-s(\mu)] d\mu ,\quad
s_0(\eta) = 4\eta^2,\quad G(\eta, \mu)= \frac 1 \eta \ln \left|\frac{\eta+\mu}{\eta-\mu} \right|\, ,
\end{equation}  
where $\Gamma \subset [0,\infty)$.

\section{Soliton condensate}
\label{sec:cond}

The density of states (DOS) of the soliton condensate and the corresponding flux solve the nonlinear dispersion relations (NDRs):
 \begin{align}
    \label{eq:condensate}
       & \int_{\Gamma_N} G(\eta,\mu) f^{(N)}(\mu) d\mu = 1,\quad \int_{\Gamma_N} G(\eta,\mu) g^{(N)}(\mu) d\mu = s_0(\eta),\quad \forall \eta \in \Gamma_N,
\end{align}
where $g^{(N)}(\mu) = s^{(N)}(\mu) f^{(N)}(\mu)$ and $\Gamma_N = [0,\beta_1) \cup (\beta_2,\beta_3) \dots \cup (\beta_{2N},\beta_{2N+1})$; $N$ is the condensate genus. The solution $f^{(N)}$ and $s^{(N)}$ for an arbitrary genus $N$ are given in \cite{congy_dispersive_2023}. By convention, all the quantities denoted with the upper index $(N)$ depend explicitly on the $\beta_i$'s.

We first focus on the case $N=0$ with $\Gamma_0=[0,\beta)$, where we drop the index $1$ in $\beta_1$ for convenience. If we take the odd extension of $f^{(0)}(\mu)$ and $g^{(0)}(\mu)$, and multiply the two equations in \eqref{eq:condensate} by $\eta$, we obtain:
 \begin{align}
-\int_{-\beta}^{\beta} \ln|\eta-\mu| f^{(0)}(\mu) d\mu = \eta,\quad -\int_{-\beta}^{\beta} \ln|\eta-\mu| g^{(0)}(\mu) d\mu = 4\eta^3,
\end{align}
which yields after differentiation with respect to $\eta$ the principal value integrals
\begin{equation}
    \label{eq:fht_zero}
    \pval \int_{-\beta}^{\beta} \frac{f^{(0)}(\mu)}{\mu-\eta} d\mu = 1,\quad
    \pval \int_{-\beta}^{\beta} \frac{g^{(0)}(\mu)}{\mu-\eta} d\mu = 12 \eta^2.
\end{equation}
One can check by direct substitution that (see also \cite{congy_dispersive_2023})
\begin{equation}
\label{eq:genus0}
f^{(0)}(\eta) = \frac{\eta}{\pi\sqrt{\beta^2-\eta^2}},\quad g^{(0)}(\eta)  = \frac{\eta(12\eta^2-6\beta^2)}{\pi\sqrt{\beta^2-\eta^2}},\quad \forall \eta \in [0,\beta).
\end{equation}
Let $\beta=\beta(x,t)$, then the substitution of \eqref{eq:genus0} in \eqref{eq:kin} yields the modulation equation:
\begin{equation}
    \label{eq:beta0a}
    \beta_t+ v^{(0)} \beta_x=0,\quad v^{(0)}=6\beta^2.
\end{equation}
We also have the useful identities for $\eta>\beta$:
\begin{equation}
\label{eq:tracer2}
  \begin{split}
    &\int_0^{\beta} G(\eta,\mu) f^{(0)}(\mu) d\mu =
    1-\sqrt{1-\frac{\beta^2}{\eta^2}},\\
    &\int_0^{\beta} G(\eta,\mu) g^{(0)}(\mu) d\mu =
    4\eta^2 - (4\eta^2+2\beta^2)\sqrt{1-\frac{\beta^2}{\eta^2}}.
  \end{split}
\end{equation}

\section{Composite soliton gas}

Let $\Gamma = \Gamma_N \cup \tilde \Gamma$, $\Gamma_N \cap \tilde \Gamma= \emptyset$ and make the ansatz
\begin{equation}\label{eq:compositea}
f(\eta) = f^{(N)}(\eta) + \ft(\eta) - \int_{\Gammat} p^{(N)}(\eta,\mu) \ft(\mu) d\mu,
\end{equation}
with $\ft(\eta)$ supported on $\Gammat$ and equaling zero at the endpoints of $\Gammat$, and $p^{(N)}(\eta,\mu)$ supported on $\Gamma_N \times \Gammat$. The latter distribution is uniquely selected by solving the integral equation
 \begin{equation}
    \label{eq:induceda}
    \begin{split}
\int_{\Gamma_N} G(\eta,z) p^{(N)}(z,\mu) dz = G(\eta,\mu),\quad \forall (\eta,\mu) \in \Gamma_N \times \Gammat,
        \end{split}
    \end{equation} 
such that
\begin{equation}\label{eq:cond_compa}
    \int_{\Gamma} G(\eta,\mu) f(\mu) d\mu = 1,\quad \forall \eta \in \Gamma_N.
\end{equation}

\subsection{Dressed equation of state}

Let's verify that $s(\eta)$ defined by
\begin{equation}
\label{eq:ga}
    g(\eta)=s(\eta) f(\eta)  = g^{(N)}(\eta) + \gt(\eta) - \int_{\Gammat} p^{(N)}(\eta,\mu) \gt(\mu) d\mu,
\end{equation}
where $\gt(\eta)=\st(\eta)\ft(\eta)$ is supported on $\Gammat$, solves the equation of state \eqref{eq:statea} if $\st(\eta)$ solves the ``dressed equation of state'' defined later in \eqref{eq:stateta}. 
Since $p^{(N)}(\eta,\mu)$ solves \eqref{eq:induceda}, we have
\begin{equation}\label{eq:cond_comp2}
    \int_{\Gamma} G(\eta,\mu) g(\mu) d\mu = s_0(\eta),\quad \forall \eta \in \Gamma_N.
\end{equation}
The equation of state \eqref{eq:statea} can be written in the following form
\begin{equation}
s(\eta) \left( 1- \int_\Gamma G(\eta, \mu) f(\mu) d\mu\right)= s_0(\eta) - \int_\Gamma G(\eta, \mu) g(\mu) d\mu.
\end{equation}  
For $\eta\in\Gamma_N$ the left hand side cancels using \eqref{eq:cond_compa} and the right hand side using \eqref{eq:cond_comp2}, implying that the equation of state is trivially satisfied for $\eta\in\Gamma_N$.

Substituting \eqref{eq:compositea} and \eqref{eq:ga} in the equation of state \eqref{eq:statea}, we have for $\eta \in \Gammat$
\begin{equation}
\begin{split}
    \st(\eta) &= s_0(\eta) + \int_{\Gamma_N} G(\eta,\mu) f^{(N)}(\mu) \st(\eta) d\mu - \int_{\Gamma_N} G(\eta,\mu) g^{(N)}(\mu) d\mu\\
    &- \int_{\Gamma_N} G(\eta,z) \left(\int_{\Gammat} p^{(N)}(z,\mu) \ft(\mu)[\st(\eta)- \st(\mu)] d\mu \right) dz + \int_{\Gammat} G(\eta,\mu) \ft(\mu)[\st(\eta)- \st(\mu)] d\mu.
\end{split}
\end{equation}
Swapping the order of integration in the double integral, we obtain
\begin{equation}
\begin{split}
    \left(1-\int_{\Gamma_N} G(\eta,\mu) f^{(N)}(\mu) d\mu\right)\st(\eta) = s_0(\eta) - \int_{\Gamma_N} G(\eta,\mu) g^{(N)}(\mu) d\mu\\
     + \int_{\Gammat} \left(G(\eta,\mu)-\int_{\Gamma_N} G(\eta,z) p^{(N)}(z,\mu) dz\right) \ft(\mu)[\st(\eta)- \st(\mu)] d\mu,
\end{split}
\end{equation}
which can be written
\begin{align}
&\label{eq:stateta}
    \st(\eta) = \st_0(\eta) + \int_\Gammat \Gt(\eta,\mu) \ft(\mu)[\st(\eta)-\st(\mu)] d\mu,\\
    \label{eq:s0t}
    &\st_0(\eta)= \frac{s_0(\eta)- \int_{\Gamma_N} G(\eta,\mu) g^{(N)}(\mu) d\mu}{1- \int_{\Gamma_N} G(\eta,\mu) 
    f^{(N)}(\mu) d\mu},\\
    \label{eq:Gt}
&\Gt(\eta,\mu) = \frac{G(\eta,\mu)- \int_{\Gamma_N} G(\eta,z) p^{(N)}(z,\mu) dz}{1- \int_{\Gamma_N} G(\eta,z) 
    f^{(N)}(z) dz}.
\end{align}

\subsection{Dressed kinetic equation}

Suppose now that $\ft(\eta)=\ft(\eta;x,t)$, $\gt(\eta)=\gt(\eta;x,t)$ and $\beta_i=\beta_i(x,t)$. \eqref{eq:kin} trivially reads for $\eta \in \Gammat$:
\begin{equation}
\label{eq:kinta}
    \ft_t+\gt_x=0.
\end{equation}
Substituting \eqref{eq:compositea} and \eqref{eq:ga} in \eqref{eq:kin} and using the new kinetic equation \eqref{eq:kinta}, we obtain
\begin{equation}
\label{eq:newmod}
\begin{split}
    \sum_{j=1}^{2N+1} &\left(\partial_{\beta_j} f^{(N)}(\eta) - \int_\Gammat \partial_{\beta_j} p^{(N)}(\eta,\mu)\ft(\mu) d\mu\right) \beta_{j,t} \\
    + &\left(\partial_{\beta_j} g^{(N)}(\eta) - \int_\Gammat \partial_{\beta_j} p^{(N)}(\eta,\mu)\gt(\mu) d\mu\right) \beta_{j,x} = 0.
\end{split}
\end{equation}
Fix $i \in \{1,\ldots,2N+1\}$ and let $F^{(N)}(\eta) = \int f^{(N)}(\eta) d\eta$ with integration constant chosen so that $F^{(N)}(\beta_i) = 0$. 
Integrating the first equation  by parts in \eqref{eq:condensate} we obtain:
\begin{equation}
    \sum_{j\neq i} (-1)^{j+1} G(\eta,\beta_j) F^{(N)}(\beta_j) - \int_{\Gamma_N} \partial_\mu G(\eta,\mu) F^{(N)}(\mu) d\mu = 1,\quad \forall \eta \in \Gamma_N.
\end{equation}
Differentiating with respect to $\beta_i$, we obtain a linear equation for $\partial_{\beta_i}F^{(N)}(\mu)$:
\begin{equation}
    \sum_{j\neq i} (-1)^{j+1} G(\eta,\beta_j) \partial_{\beta_i}F^{(N)}(\beta_j) - \int_{\Gamma_N} \partial_\mu G(\eta,\mu) \partial_{\beta_i}F^{(N)}(\mu) d\mu = 0,\quad \forall \eta \in \Gamma_N
\end{equation}
Similarly, let $P^{(N)}(\eta,\mu) = \int p^{(N)}(\eta,\mu) d\eta$ with integration constant chosen so that $P^{(N)}(\beta_i,\mu) = 0$, and integrate eq.~\eqref{eq:induceda}  by parts.  Then, differentiating with respect to $\beta_i$, we obtain the same linear equation for $\partial_{\beta_i}P^{(N)}(\eta,\mu)$:
\begin{equation}
    \sum_{j\neq i} (-1)^{j+1} G(\eta,\beta_j) \partial_{\beta_i}P^{(N)}(\beta_j,\mu) - \int_{\Gamma_N} \partial_z G(\eta,z) \partial_{\beta_i}P^{(N)}(z,\mu) dz = 0,\quad \forall \eta \in \Gamma_N.
\end{equation}
By uniqueness of the solution to this linear integral equation, $\partial_{\beta_i}F^{(N)}(\eta)$ and $\partial_{\beta_i}P^{(N)}(\eta,\mu)$ are linearly dependent
\begin{equation}
\label{eq:a}
    \partial_{\beta_i}P^{(N)}(\eta,\mu) = a_i^{(N)}(\mu) \partial_{\beta_i}F^{(N)}(\eta) \; \Rightarrow \; \partial_{\beta_i}p^{(N)}(\eta,\mu) = a_i^{(N)}(\mu) \partial_{\beta_i}f^{(N)}(\eta).
\end{equation}
With a similar proof, we can show that
\begin{equation}
    \partial_{\beta_i} g^{(N)}(\eta) = v_i^{(N)} \partial_{\beta_i} f^{(N)}(\eta),
\end{equation}
where $v_i^{(N)}$ is the $i^{\rm th}$ characteristic velocity of the Whitham modulation system; see \cite{congy_dispersive_2023}.

Thus \eqref{eq:newmod} reduces to the ``dressed'' modulation equations
\begin{equation}
\label{eq:Whithama}
    \beta_{i,t} + \gamma_i \beta_{i,x} = 0,\quad \gamma_i=\frac{\displaystyle v_i^{(N)} - \int_\Gammat a_i^{(N)}(\mu)\gt(\mu) d\mu}{\displaystyle 1 - \int_\Gammat a_i^{(N)}(\mu)\ft(\mu) d\mu},\quad a_i^{(N)}(\mu)=\frac{\partial_{\beta_i}p^{(N)}(\eta,\mu)}{\partial_{\beta_i}f^{(N)}(\eta)}.
\end{equation}

\section{Interaction with a genus 0 condensate}

\subsection{Derivation of dressed coefficients}

Let $N=0$.
If we take the odd continuation of $p^{(0)}(z,\mu)$ with respect to $z$ and apply the same procedure outlined in Section~\ref{sec:cond}, \eqref{eq:induceda} reads
\begin{equation}
   \pval \int_{-\beta}^{\beta} \frac{p^{(0)}(z,\mu)}{z-\eta} dz = \frac{2\mu}{\mu^2-\eta^2}.
\end{equation}
By direct substitution, one can verify that 
\begin{equation}\label{eq:N0}
p^{(0)}(\eta,\mu)=\frac{2\eta\sqrt{\mu^2-\beta^2} }{\pi(\mu^2-\eta^2)\sqrt{\beta^2-\eta^2}},\quad \forall (\eta,\mu) \in [0,\beta) \times \Gammat.
\end{equation}
We also have the identity for $\eta>\beta$ and $\mu>\beta$:
\begin{equation}\label{eq:lim2}
    \int_0^{\beta}G(\eta,z) p^{(0)}(z,\mu)dz =
    \frac{1}{\eta}\ln \left|\frac{\eta+\mu}{\eta-\mu} \right|-\frac{1}{\eta}\ln\left| \frac{\sqrt{\eta^2-\beta^2}+\sqrt{\mu^2-\beta^2}}{\sqrt{\eta^2-\beta^2}-\sqrt{\mu^2-\beta^2}}\right|.
\end{equation}
Substituting \eqref{eq:tracer2}, \eqref{eq:lim2} in the coefficient formulae \eqref{eq:s0t}, \eqref{eq:Gt}, \eqref{eq:Whithama}, we obtain for $\eta>\beta$ and $\mu>\beta$:
\begin{align}
\label{eq:case0}
&\st_0(\eta) = 2\beta^2+4\eta^2,\\ 
&\Gt(\eta,\mu)=\frac{1}{\sqrt{\eta^2-\beta^2}}\ln\left| \frac{\sqrt{\eta^2-\beta^2}+\sqrt{\mu^2-\beta^2}}{\sqrt{\eta^2-\beta^2}-\sqrt{\mu^2-\beta^2}}\right|= G\Big(\sqrt{\eta^2-\beta^2},\sqrt{\mu^2-\beta^2}\Big),\\
&a^{(0)}(\eta)  = \frac{2}{\sqrt{\eta^2-\beta^2}}.
\end{align}

\subsection{Hydrodynamic reduction}

Inserting
\begin{equation}\label{polya}
\ft(\eta;x,t)= \sum \limits_{j=1}^M w_j(x,t) \delta(\eta - \xi_j),
\end{equation}
into eqs.~\eqref{eq:stateta} and \eqref{eq:kinta}, we obtain 
\begin{equation}\label{hyd_consa}
\begin{split}
& w_{j,t}+(c_j\,w_j)_x=0, \quad c_j = 2\beta^2+ 4\xi_j^2 +  \sum_{k\neq j} \Gt_{jk} w_k(c_j-c_k), \quad \ j=1, \dots, M,
\end{split}
\end{equation}
where $\Gt_{jk} =\Gt (\xi_j, \xi_k)$ and the $w_j$ are
coupled to $\beta$ through \eqref{eq:Whithama} with \eqref{eq:N0}
\begin{equation}
\label{hyd_cons_laa}
\begin{split}
 &\beta_{t}+ \gamma \beta_{x}=0,\quad \gamma = \frac{\displaystyle 6\beta^2 - \sum_{j=1}^M\frac{2c_jw_j}{\sqrt{\xi_j^2-\beta^2}}}{\displaystyle1-\sum_{j=1}^M \frac{2w_j}{\sqrt{\xi_j^2-\beta^2}}}.
\end{split}
\end{equation}

By direct substitution, one can verify that the transformation
\begin{equation}
    \label{eq:qj_defn}
    q_j =  \left ( 1-\sum_{k\neq j} \Gt_{jk} w_k \right ) \sqrt{\xi_j^2-\beta^2}/w_j - \ln ( 1-\beta^2/\xi_j^2 ), \quad j = 1, \ldots, M,
\end{equation}
diagonalizes eq.~\eqref{hyd_consa} as
\begin{equation}\label{eq:qja}
    q_{j,t}+ c_j q_{j,x} = 0,\quad j=1,\dots,M.
\end{equation}

We now prove that the quasi-linear system \eqref{eq:qja} is linearly degenerate. Let's define the matrix $\Ht$ with
\begin{equation}
    \Ht_{jk} = \begin{cases}
        \sqrt{\xi_j^2-\beta^2} \Gt_{jk}, & k \neq j,\\
        q_j+ \ln \left( 1-\frac{\beta^2}{\xi_j^2} \right), & k =j.
    \end{cases}
\end{equation}
such that $w_j$ can be written in term of $\beta$ and $q_j$ (see \eqref{eq:qj_defn}):
\begin{equation}
\label{eq:sigmatb}
    w_j = \sum_{k=1}^M \big(\Ht^{-1}\big)_{jk} \sqrt{\xi_k^2-\beta^2}.
\end{equation}
Similarly, combining \eqref{hyd_consa} and \eqref{eq:qj_defn}, we obtain
\begin{equation}
    c_jw_j = \sum_{k=1}^M \big(\Ht^{-1}\big)_{jk} ( 2\beta^2+ 4\xi_k^2 ) \sqrt{\xi_k^2-\beta^2}.
\end{equation}
Differentiating \eqref{eq:sigmatb} with respect to $q_j$ and using the identity
\begin{equation}
    \frac{\partial \big(\Ht^{-1}\big)_{jk}}{\partial q_j} = 
    \frac{\partial \big(\Ht^{-1}\big)_{jk}}{\partial \Ht_{jj}} = \sum_{l,m-1}^M -\big(\Ht^{-1}\big)_{jl} \frac{\partial \Ht_{lm}}{\partial \Ht_{jj}} \big(\Ht^{-1}\big)_{mk} =-\big(\Ht^{-1}\big)_{jj}\big(\Ht^{-1}\big)_{jk},
\end{equation}
we obtain
\begin{equation}
    \frac{\partial w_j}{\partial q_j} = -\big(\Ht^{-1}\big)_{jj} w_j,\quad \frac{\partial (c_j w_j)}{\partial q_j} = -\big(\Ht^{-1}\big)_{jj} c_j w_j,
\end{equation}
which implies the linear degeneracy criterion
\begin{equation}
    \frac{\partial c_j}{\partial q_j} = \frac{1}{w_j} \left( \frac{\partial c_j w_j}{\partial q_j} - c_j \frac{\partial w_j}{\partial q_j}\right) = 0.
\end{equation}

\subsection{Bi-chromatic soliton gas: explicit expressions and integrability}

Let $M=2$. We can express $w_j$ and $c_j$ explicitly in terms of $q_j$ and $\beta$:
\begin{align}
\label{eq:case2}
\begin{split}
    &w_j=\frac{\sqrt{\xi_k^2-\beta^2}h-\sqrt{\xi_j^2-\beta^2}\left[q_k+\ln \left( 1-\frac{\beta^2}{\xi_k^2} \right)\right]}
    {h^2-\left[q_j+\ln \left( 1-\frac{\beta^2}{\xi_j^2} \right)\right]\left[q_k+\ln \left( 1-\frac{\beta^2}{\xi_k^2} \right)\right]},\\
    &c_j = 4\xi_j^2+2\beta^2+\frac{4h (\xi_j^2-\xi_k^2)\sqrt{\xi_k^2-\beta^2}}{\sqrt{\xi_j^2-\beta^2}\left[q_k+\ln \left( 1-\frac{\beta^2}{\xi_k^2} \right)\right]-\sqrt{\xi_k^2-\beta^2} h},\\
    &h = \Ht_{12} = \Ht_{21} = \ln\left| \frac{\sqrt{\xi_1^2-\beta^2}+\sqrt{\xi_2^2-\beta^2}}{\sqrt{\xi_1^2-\beta^2}-\sqrt{\xi_2^2-\beta^2}}\right|,
    \end{split}
\end{align}
with $(j,k) \in \{ (1,2), (2,1)\}$. We obtain an explicit expression of $\gamma$ in terms of $(\beta,q_1,q_2)$ by substituting \eqref{eq:case2} in \eqref{hyd_cons_laa}. We can notably show that
\begin{equation}
\begin{split}
    &\partial_\beta \left( \frac{\partial_{q_2}c_1}{c_2-c_1}\right) - \partial_{q_2} \left( \frac{\partial_{\beta}c_1}{\gamma-c_1}\right) = 0,\\
    &\partial_\beta \left( \frac{\partial_{q_1}c_2}{c_1-c_2}\right) - \partial_{q_1} \left( \frac{\partial_{\beta}c_2}{\gamma-c_2}\right) = 0,\\
    &\partial_{q_2} \left( \frac{\partial_{q_1}\gamma}{c_1-\gamma}\right) - \partial_{q_1} \left( \frac{\partial_{q_2}\gamma}{c_2-\gamma}\right) = 0.
\end{split}
\end{equation}
Thus the diagonal system \eqref{hyd_cons_laa}, \eqref{eq:qja} for $M=2$ is semi-Hamiltonian and therefore integrable by the generalized hodograph transform \cite{tsarev_poisson_1985, tsarev_geometry_1991}.

\section{Numerical methods}

We investigate in this work the Riemann problem for the kinetic equation \eqref{eq:kin}, \eqref{eq:statea} with the step initial condition:
\begin{equation}
\label{eq:inita}
    f(\eta;x,t=0) = 
    \begin{cases}
        w_1^0 \delta(\eta-\xi_1) + w_2^0 \delta(\eta-\xi_2), &x<0,\\
        f^{(0)}(\eta) \text{ with } \beta=1, &x>0,
    \end{cases}
\end{equation}
i.e. a bi-chromatic soliton gas for $x<0$ and a pure condensate at $x>0$.
This DOS reduces for the hydrodynamic system \eqref{hyd_consa}, \eqref{hyd_cons_laa} to the initial condition 
\begin{equation}
    (\beta,w_1,w_2) = \begin{cases}
        (0,w_{1}^0,w_{2}^0), &x<0,\\
        (1,0,0), &x>0.
    \end{cases}
\end{equation}
See eq.~(19) in the main text.

In the numerical simulations, realizations of the SG with the initial DOS \eqref{eq:inita} are implemented with the initial data
\begin{align}
\label{eq:u0}
    u(x,0) = u_n(x,0) + \overline{u}_0(x),\quad
    \overline{u}_0(x) = \frac{1}{2}+\frac{1}{2} \tanh \left(\frac{x}{\Delta x} \right).
\end{align}
Here, $u_n$ is the $n$-soliton solution of the KdV equation with $n \gg 1$ parameterized by the spectral parameters $(\eta_1,\dots,\eta_n)$ and position phases $(x^0_1,\dots,x^0_n)$. The spectral parameters and position phases are randomly distributed to model the bi-chromatic SG at $x<0$. The smooth function $\overline{u}_0(x)$ models the initial mean flow at $x>0$.  Note that some non-solitonic (continuous) spectrum develops if $\Delta x$ is too small leading to small amplitude (radiation) waves. We use $\Delta x =5$.
We evolve the SG numerically by first preparing $R=100$ initial conditions \eqref{eq:u0} with $n=200$ solitons each, and then solving the corresponding  KdV initial value problem. The KdV equation is solved using a pseudospectral-Fourier approximation of spatial derivatives and a 4th order Runge-Kutta timestepping method, see \cite{trefethen_spectral_2000}. This implies that we are utilizing periodic boundary conditions so we choose the domain boundaries far from the initial, smoothed box $\overline{u}_0(x)$ edges.  We then average the different solutions at $t=65$ to obtain the variation of $\langle u \rangle$ and $\langle u^2 \rangle$.

The algorithm to numerically compute the $n$-soliton solution is described in Sec.~\ref{sec:app_algo}. The distributions of the spectral parameters and phases for the bi-chromatic SG are given in Sec.~\ref{sec:bisg}. Finally, the numerical averaging procedure is explained in Sec.~\ref{sec:aver}. 

\subsection{Algorithm for the $n$-soliton solution }
\label{sec:app_algo}

The algorithm generating the exact $n$-soliton of the KdV equation via the Darboux transformation was originally presented in \cite{huang1992darboux}.  Details for the implementation of this algorithm within the context of KdV SGs can be found in \cite{congy_riemann_2025}. The numerical scheme suffers from significant round-off error during the summation of exponentially small and large
values when the number of solitons is larger than $n = 10$.  To deal with this, we use a multi-precision routine following the method introduced in 
\cite{gelash2018strongly}.
By construction, the algorithm yields the $n$-soliton solution of \eqref{eq:kdv} if $n$ is even and its equivalent form 
$u_t-6uu_x+u_{xxx}=0$
obtained from  \eqref{eq:kdv} by the reflection $u \to - u$ if $n$ is odd.
Let $J_n,\bar{J}_n\in \mathbb{R}^{2 \times 2}$ be the Jost functions defined recursively by the Darboux transformations $D(\eta)$ and $\overline{D}(\eta)$:
\begin{equation}
\begin{split}
    J_{n}(z)=D_{n}(z) J_{n-1}(z),\quad D_{n}(z)=I+\frac{2 \eta_{n}}{z- \eta_{n}} P_{n},\\
    \bar{J}_{n}(z)=\overline{D}_{n}(z) \bar{J}_{n-1}(z), \quad \overline{D}_{n}(z)=I-\frac{2\eta_{n}}{z+\eta_{n}} \overline{P}_{n},
\end{split}
\end{equation}
where $P_n$ and $\overline{P}_n$ are the projection matrices
\begin{equation}
  P_{n}=\sigma_2\overline{P}^{ \textrm{T}}_{n}\sigma_2=\frac{J_{n-1}\left(-\eta_{n}\right)\left(\begin{array}{c}
        -b_n \\
        1
      \end{array}\right)\left(\begin{array}{ll}
        b_n & 1
      \end{array}\right) \bar{J}_{n-1}^{-1}\left(\eta_{n}\right)}{\left(\begin{array}{ll}
        b_n & 1
      \end{array}\right) \bar{J}_{n-1}^{-1}\left(\eta_{n}\right) J_{n-1}\left(-\eta_{n}\right)\left(\begin{array}{c}
        -b_n \\
        1
      \end{array}\right)},\quad b_n=\left(-1\right)^{n}\exp\left(2 \eta_n x^{0}_{n}\right).
\end{equation}
The Jost solutions for the initial $0$-soliton solution are
\begin{equation}
  J_{0}(z)=\bar{J}_{0}(z)=\left(\begin{array}{cc}
      \exp \left[z x-4 z^{3} t\right] & -\exp \left[-z x+4 z^{3} t\right]         \\
      0                                     & - 2 z \exp \left[-z x+4 z^{3} t\right]
    \end{array}\right),
\end{equation}
and the $n$-soliton solution is given recursively by
\begin{equation}
\label{eq:un}
  u_{n}(x,t)=u_{n-1}(x,t)+4 \eta_{n}\left(P_{n}\right)_{21}.
\end{equation}

\subsection{Bi-chromatic SG}
\label{sec:bisg}

\begin{figure}
    \centering
    \includegraphics[scale=0.5]{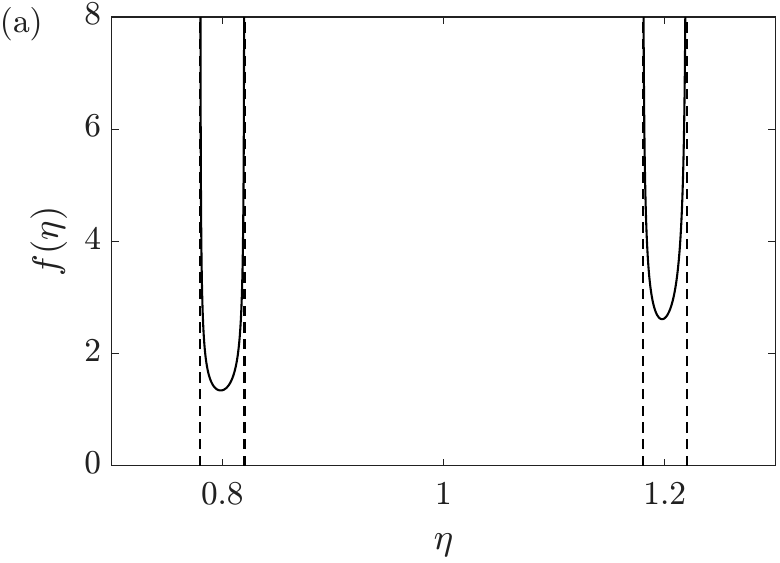}\hspace{1cm}
    \includegraphics[scale=0.5]{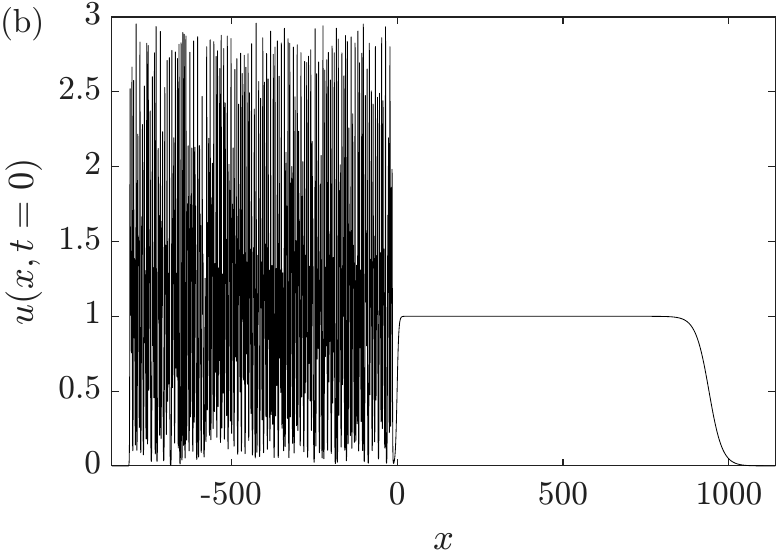}
    \caption{(a) The initial DOS \eqref{eq:dilute} modeling a bi-chromatic SG with $\xi_1=0.8$ and $\xi_2=1.2$; the dashed lines indicate the endpoints of the support $\Gamma_2 = (\xi_1-\varepsilon_1,\xi_1+\varepsilon_1) \cup (\xi_2-\varepsilon_2,\xi_2+\varepsilon_2)$. (b) Numerical implementation of the initial condition \eqref{eq:u0}.}
    \label{fig:example2}
\end{figure}

The spectral parameters $\eta_k$ in the $n$-soliton \eqref{eq:un} are distinct and the idealized DOS $f(\eta)=w_1^0 \delta(\eta-\xi_1) + w_2^0 \delta(\eta-\xi_2)$ for the bi-chromatic soliton gas must be approximated by a continuous distribution. Following \cite{congy_riemann_2025}, the delta-function distribution is implemented via the ``dilute'' condensate DOS:
\begin{equation}
\label{eq:dilute}
    f(\eta)= C f^{(2)}(\eta) \quad \text{with:} \quad \beta_1=0,\; \beta_{2j} = \xi_j-\varepsilon_j,\;\beta_{2j+1} = \xi_j+\varepsilon_j,\; 0<\varepsilon_j,\;C<1,\;j\in\{1,2\}.
\end{equation}
The support of this DOS is $\Gamma_2 = (\xi_1-\varepsilon_1,\xi_1+\varepsilon_1) \cup (\xi_2-\varepsilon_2,\xi_2+\varepsilon_2)$ where we choose $\varepsilon_j \ll 1$ such that the spectral parameters are distributed in a small neighborhood of $\xi_j$. The component densities $w_j^0$ are prescribed by
\begin{equation}
    w_j^0 = \int_{\beta_{2j}}^{\beta_{2j+1}} C f^{(2)}(\eta) d\eta,
\end{equation}
and the choice of parameters $(C,\varepsilon_1,\varepsilon_2)$ fixes the component densities $w_j^0$. We choose $\varepsilon_1=\varepsilon_2=0.02$, which yields the component densities
\begin{equation}
\label{eq:densities}
\begin{split}
        &(\xi_1,\xi_2)=(1.2,1.3),\quad (w_1^0,w_2^0) = (0.108,0.141),\\
    &(\xi_1,\xi_2)=(0.8,1.2),\quad (w_1^0,w_2^0) = (0.084,0.165).
\end{split}
\end{equation}
The variation of $f(\eta)$ given by \eqref{eq:dilute} for the second set of parameters is displayed in Fig.~\ref{fig:example2}.

The spectral parameters $\eta_k$ are then distributed using the relation
\begin{equation}
    \int^{\eta_k} \phi(\eta) d\eta = \frac{k}{n},\quad k=1,\dots,n,
\end{equation}
where $\phi(\eta)$ is the normalized DOS:
\begin{equation}
    \phi(\eta) = \frac{f^{(2)}(\eta)}{\kappa^{(2)}},\quad \kappa^{(2)} = \int_{\Gamma_2} f^{(2)}(\eta) d \eta.
\end{equation}
The SG of DOS \eqref{eq:dilute} is not a soliton condensate since $C \neq 1$ and the position phases $x_k^0$ are randomly distributed, i.e. the $n$-soliton solution implementing the SG realization is stochastic. $x_k^0$ are uniformly distributed in the interval:
\begin{equation}
    I = \left[-\ell + \frac{C}{2} \ell,\frac{-C}{2} \ell\right],\quad
    \ell = \frac{n}{C \kappa^{(2)}},
\end{equation}
such that the $n$-soliton solution $u_n(x,t=0)$ implements the uniform SG of DOS \eqref{eq:dilute} in the spatial region $x\in [-\ell,0]$, see \cite{congy_riemann_2025}; $u_n(x,t=0)$ decays exponentially fast outside the region $[-\ell,0]$. An example numerical implementation of the initial condition \eqref{eq:u0} is shown in Fig.~\ref{fig:example2}.

\subsection{Averaging procedure}
\label{sec:aver}

On a sufficiently large spatial scale $L$ (much larger than the typical soliton width), the nonlinear wave field in a SG is an ergodic random process \cite{el_soliton_2021}. The ergodicity property implies that ensemble
averages can be replaced by spatial averages. In order to accelerate the calculation of statistical moments
$\left\langle u  \right\rangle$ are obtained by averaging over the ensemble and spatially
\begin{equation}
	\label{eq:aver_num}
	\left\langle u(x,t)^m  \right\rangle = \frac{1}{L} \int_{x-L/2}^{x+L/2}\left( \frac{1}{R}\sum_{i=1}^R u(x,t)^m \right) d x, \quad m = 1, 2,
\end{equation}
where $u(x,t)$ is the numerical solution of the KdV equation \eqref{eq:kdv} for different samples of the random initial condition \eqref{eq:u0}.
Since the components densities are $w_j^0 = O(10^{-1})$, see \eqref{eq:densities}, we choose the spatial scale $L=1/10^{-1}=10$ corresponding to the typical spacing between solitons.

\end{document}